\documentclass[sigconf]{acmart}
\AtBeginDocument{%
	}

\usepackage[ruled,linesnumbered]{algorithm2e}
\usepackage[caption=false,textfont=sf]{subfig}

\acmConference[ISACom]{ACM MobiCom Workshop on Integrated Sensing and Communication Systems}{24-28 October,
	2022}{Sydney, Australia}

\begin{document}

	\title{Iterative Sparse Recovery based Passive Localization in Perceptive Mobile Networks}

	\author{Lei~Xie and Shenghui Song}
	\affiliation{%
		\institution{Dept. of ECE, The Hong Kong University of Science and Technology, Hong Kong}
	}
	\email{Email: {eelxie, eeshsong}@ust.hk}

	\begin{abstract}
		Perceptive mobile networks (PMNs) were proposed to integrate sensing capability into current cellular networks where multiple sensing nodes (SNs) can collaboratively sense the same targets. Besides the active sensing in traditional radar systems, passive sensing based on the uplink communication signals from mobile user equipment may play a more important role in PMNs, especially for targets with weak electromagnetic wave reflection, e.g., pedestrians. However, without the properly designed active sensing waveform, passive sensing normally suffers from low signal to noise power ratio (SNR). As a result, most existing methods require a large number of data samples to achieve an accurate estimate of the covariance matrix for the received signals, based on which a power spectrum is constructed for localization purposes. Such a requirement will create heavy communication workload for PMNs because the data samples need to be transferred over the network for collaborative sensing. To tackle this issue, in this paper we leverage the sparse structure of the localization problem to reduce the searching space and propose an iterative sparse recovery (ISR) algorithm that estimates the covariance matrix and the power spectrum in an iterative manner. Experiment results show that, with very few samples in the low SNR regime, the ISR algorithm can achieve much better localization performance than existing methods.
	\end{abstract}
	
	
	\maketitle
	
	\section{Introduction}
	With the development of innovative applications such as autonomous driving, sensing becomes an important service for future wireless networks. To this end, the recently proposed integrated sensing and communication (ISAC) provides a promising platform to exploit the synergy between sensing and communication. As a special type of ISAC system, perceptive mobile networks (PMNs) were proposed to add sensing capability to cellular networks \cite{xie2021perceptive,xie2022networked,xie2022collaborative,9296833} without interfering with the communication service. The adoption of millimeter wave (mmWave) by 5G and beyond systems further enables hardware and software reuse between sensing and communication.  
	
	Given the existing uplink communication from the user equipment (UEs) to the base stations, passive sensing, which utilizes uplink signals for sensing purposes instead of actively transmitting probing signals, demonstrates its advantage for targets with weak electromagnetic wave reflection, e.g., pedestrians. Furthermore, without the need to transmit sensing signals, passive sensing is more energy-efficient and causes less latency than active sensing. However, without the well-designed waveform in active sensing, the received signal to noise power ratio (SNR) for passive sensing is normally low. The situation is even worse in the mmWave band due to the very high pathloss and omni-directional uplink signals. 
	
	There are two types of localization methods for passive sensing. The most popular one is the indirect localization method which has two steps. In the first step, some intermediate parameters of the targets, e.g., distance, time-difference-of-arrival (TDOA), and angle of arrival (AOA) are estimated. In the second step, the localization of the target is achieved based on the intermediate parameters.  A closed-form localization solution based on weighted least squares estimation was proposed in \cite{noroozi2015target}. The authors of \cite{noroozi2017iterative} modelled the localization based on intermediate parameters as a quadratically constrained quadratic programming (QCQP) problem, which is non-convex and thus hard to solve. To address this issue, the QCQP problem is reformulated as a linearly constrained quadratic programming (LCQP) problem, and solved by the iterative constrained weighted least squares method. On the other hand, the localization performance also suffers from outlier samples \cite{9686699}.  To handle this issue, \cite{liang2016robust} proposed a robust localization method by exploiting the maximum correntropy criterion. The authors of \cite{liu2020radar} investigated predictive beamforming to track vehicles by exploiting the dual-functional radar-communication (DFRC) technique, where the positions of the vehicles are recovered based on time delay and Doppler frequency. However, the indirect methods may lose information when estimating intermediate parameters in the first step, which causes poor localization performance especially with low SNR. Furthermore, there is an intrinsic issue for indirect localization methods with multiple targets, i.e., the matching of the intermediate parameters with the locations of different targets \cite{tzoreff2017expectation}. In particular, even if the intermediate parameters are perfectly estimated, the final localization results may be wrong due to the mismatching issues. 
	
	Another class of passive localization methods estimates the target positions directly without the intermediate parameters estimation, and is called direct localization. Compared with the indirect methods, direct localization is implemented at the signal level and can achieve much better performance, especially with low SNR \cite{tzoreff2017expectation}.  
	The conventional direct localization methods located the targets based on the power spectrum estimated from the sample covariance matrix (SCM) \cite{7500062,zhao2020beamspace}. The authors of \cite{7500062} proposed to locate the targets based on the minimum variance distortionless response (MVDR) method, and assumed the sample covariance matrix (SCM) to be block-diagonal. The authors of \cite{zhao2020beamspace} proposed to exploit the beam-space (BS) of the covariance matrix, but the estimation of the covariance matrix by the SCM-based method requires a large number of samples. As a result, the above two methods cannot achieve accurate localization performance when the number of samples is limited.  This situation is worse in PMNs, because collaborative sensing requires the transferring of data samples from the sensing nodes (SNs) to the central processing unit (CPU), and the large number of samples will cause high communication workload for the network. This motivates us to design a communication-efficient localization method with limited data samples.
	
	The existing direct localization methods require many samples to guarantee a satisfactory estimate of the covariance matrix of the received signals such that the power spectrum constructed based on the covariance matrix is accurate enough. To reduce the requirement on data samples, in this paper, we propose an iterative sparse recovery (ISR) method leveraging the sparse representation of the localization problem. In particular, among a large number of possible locations, only a few can hold the targets. As a result, the localization problem, i.e., the estimation of the power spectrum, does not require a large number of samples. Furthermore, by building up the connection between the covariance matrix of the received signal and the power spectrum of different locations, we propose an iterative method where the covariance matrix and the power spectrum are estimated in an iterative manner. In this way, the estimation of the covariance matrix and the construction of the power spectrum can be improved iteratively, leading to better estimation performance especially with limited samples. Compared with the existing methods, the ISR algorithm can achieve high-resolution localization performance with a few samples (even one), which is critical to reduce the communication workload of PMNs.

	\section{Preliminary}
	In this section, we present the system model of the proposed multi-target passive localization problem and introduce the existing methods in the literature.
	\subsection{Signal Model}
	Consider a PMN where $L$ SNs collaboratiely monitor $K$ targets, where $K$ is unknown\footnote{In general, the number of targets $K$ can be estimated by Akaike information criterion and Baysian information criterion \cite{1311138}.}. Depending on the network architecture, the SNs can be base stations \cite{8288677}, remote radio units (RRUs) \cite{9296833}, or target monitoring terminals (TMTs) \cite{xie2021perceptive,xie2022networked}. Assume SNs are synchronized to the same clock and are equipped with a uniform linear array (ULA) of $N_R$ antennas and the targets, e.g., UEs, only have one antenna. In this paper, we consider passive localization based on the uplink communication signals from the UEs in an orthogonal frequency division multiple (OFDM) system \cite{song2011localized}. In particular, different UEs transmit their signals at different carrier-frequencies within a given frequency band.   
	
	Let $\mathbf{u}_l$ and $\mathbf{p}_k$ denote the coordinate vectors of the $l$th SN and the $k$th target, respectively. Then, the steering vector of the $l$-th SN towards the $k$th target can be given by 
	\begin{equation}
		\mathbf{a}_l(\mathbf{p}_k)=\frac{1}{\sqrt{N_R}}\left[1,e^{j\frac{2\pi d}{\lambda}\mathbf{k}_k^T\mathbf{e}_{l}},\cdots,e^{j\frac{2\pi d}{\lambda}(N_R-1)\mathbf{k}_k^T\mathbf{e}_{l}}\right]^T,
	\end{equation}
	where $(\cdot)^T$ denotes the transpose operator, while $d$ and $\lambda$ represent the inner spacing of the ULA and the wavelength, respectively. The vector
	\begin{equation}
		\mathbf{k}_k=\frac{\mathbf{u}_l-\mathbf{p}_k}{\Vert \mathbf{u}_l-\mathbf{p}_k \Vert_2}
	\end{equation}
	denotes the direction vector from the $k$th target to the $l$th SN and $\mathbf{e}_l$ represents the unit vector parallel to the line formed by all antennas of the ULA.
	
	For ease of implementation, we do not require the SNs to know the exact carrier frequencies of all UEs, but instead they only know the central carrier and the bandwidth of the OFDM systems. 
	As a result, each SN only performs down-conversion on the received signal with the central carrier and the resulting baseband signal of the $l$-th SN is denoted by $\mathbf{y}_{l}(t)$. Then, the $l$-th SN samples its received signal and obtains
	\begin{equation}
		\label{signalmodel1}
		\mathbf{y}_{l}(n)=\sum_{k=1}^{K}\sqrt{N_R}\alpha_{l,k} \mathbf{a}_l(\mathbf{p}_k) s_{k}(n)+\mathbf{v}_l(n), n=1,2,\cdots, N_s
	\end{equation}
	where $N_s$ denotes the number of samples per symbol and the $n$th sample of the transmit signal, $s_{k}(n)$, is assumed to have zero mean and unit variance. Here, $\alpha_{l,k}$ denotes the complex channel coefficient from the $k$-th target to the $l$-th SN, which is modeled as a Gaussian random variable (RV) with zero mean and variance $\sigma_{l,k}^2$. $\mathbf{v}_l(n)$ represents the additive white Gaussian noise with covariance matrix $\sigma_v^2 \mathbf{I}$. 
	Note that synchronization between different uplink signals is not required. 
	
	It can be validated that $\mathbf{y}_{l}(n)$ follows a zero-mean Gaussian distribution with covariance matrix
	\begin{equation}\label{signallcovariance}
		\mathbf{R}_l={E}\left(\mathbf{y}_l(n)\mathbf{y}_l^H(n)\right)=\sum_{k=1}^{K}N_R\sigma_{l,k}^2 \mathbf{a}_l(\mathbf{p}_k)   \mathbf{a}_l^H(\mathbf{p}_k)+\sigma_v^2 \mathbf{I},
	\end{equation}
	where $(\cdot)^H$ denotes the conjugate transpose operator.
	By stacking the received signals of different SNs into a vector, we have
	\begin{equation}\label{signalmodel2}
		\mathbf{y}(n)=\left[\mathbf{y}_1^T(n),\mathbf{y}_2^T(n),\cdots,\mathbf{y}_L^T(n)\right]^T=\sum_{k=1}^{K}\mathbf{A}_k\mathbf{x}_k(n)+\mathbf{v}(n),
	\end{equation}
	where
	\begin{equation}\label{ak}
		\mathbf{A}_k=\left[\begin{matrix}
			\mathbf{a}_1(\mathbf{p}_k)  & \mathbf{0} & \cdots & \mathbf{0}\\
			\mathbf{0} & \mathbf{a}_2(\mathbf{p}_k)  & \cdots & \mathbf{0}\\
			\vdots & \vdots & \ddots & \vdots\\
			\mathbf{0} & \mathbf{0} & \cdots & \mathbf{a}_L(\mathbf{p}_k) \\
		\end{matrix}\right]\in \mathbb{C}^{N_R L \times L},
	\end{equation}
	denotes the steering matrix from $L$ SNs towards the $k$th target, 
	\begin{equation}\label{xkvec}
		\mathbf{x}_k(n)=\left[\alpha_{1,k},\alpha_{2,k},\cdots,\alpha_{L,k} \right]^T \sqrt{N_R} s_{k}(n)\in \mathbb{C}^{L \times 1},
	\end{equation}
	and 
	\begin{equation}\label{nvec}
		\mathbf{v}(n)=\left[\mathbf{v}_1^T(n),\mathbf{v}_2^T(n),\cdots,\mathbf{v}_L^T(n) \right]^T\in \mathbb{C}^{N_R L \times 1}.
	\end{equation}
	
	We assume that $\{\mathbf{x}_k(n)\}_{k=1}^K$ and $\mathbf{v}(n)$ are independent and statistically stationary throughout the observation period.
	Then, the covariance matrix of $\mathbf{y}(n)$ is given as
	\begin{equation}\label{signalcovariance}
		\begin{aligned}
			\mathbf{R}={E}\left(\mathbf{y}(n)\mathbf{y}(n)^H\right)
			&=\sum_{k=1}^{K}\mathbf{A}_k \mathbf{\Lambda}_k \mathbf{A}_k^H+\sigma_v^2 \mathbf{I},
		\end{aligned}
	\end{equation}
	where $\mathbf{\Lambda}_k\triangleq \mathbb{E}\left(\mathbf{x}_k(n)\mathbf{x}_k^H(n)\right)$.
	
	\subsection{Direct Localization Algorithms}
	Given the received signals at the SNs, the objective of localization is to determine the locations of the $K$ targets. For that purpose, we may discretize the concerned area into a grid of $N_G$ searching points $\{\bar{\mathbf{p}}_i\}_{i=1}^{N_G}$. The direct localization methods generate a power spectrum for the whole grid with an estimated power value for each searching point. If one target is present at one searching point, there is supposed to be a peak in the power spectrum.  
	
	Due to the difficulty to directly estimate the transmit power, one option is to estimate the received signal power from a given searching point. For that purpose, we want to collect the energy transmitted from a specific location, but suppress the interference from other locations. To this end, the MVDR method is a good candidate, because it was originally proposed as a beamforming technology to minimize the output power of interference and noise while guaranteeing a fixed gain for the desired signal \cite{1449208,xie2020recursive}. The authors of \cite{7500062} chose MVDR as the combiner and utilized the output power of the combiner as the estimated power, where the MVDR spectrum is obtained as 
	\begin{equation}\label{MVDRspec}
		\begin{split}
			P_{\text{MVDR},i}	&=\frac{1}{\sum_{l=1}^L\mathbf{a}_l^H(\bar{\mathbf{p}}_i)\widehat{\mathbf{R}}_{\text{SCM},l}^{-1}\mathbf{a}_l(\bar{\mathbf{p}}_i)},  i=1,...,N_G
		\end{split}
	\end{equation}
	with $\widehat{\mathbf{R}}_{\text{SCM},l}=\frac{1}{N_s} \sum_{n=1}^{N_s}\mathbf{y}_l(n) \mathbf{y}_l^H(n)$ representing the SCM of the $l$th SN. Note that the MVDR method estimates the covariance matrix by exploiting the block-diagonal structure
	\begin{equation}\label{SCM}
		\widehat{\mathbf{R}}_{\text{B-SCM}}=\left[\begin{matrix}
			\widehat{\mathbf{R}}_{\text{SCM},1} & \mathbf{0} & \cdots & \mathbf{0}\\
			\mathbf{0} & \widehat{\mathbf{R}}_{\text{SCM},2} & \cdots & \mathbf{0}\\
			\vdots & \vdots & \ddots & \vdots\\
			\mathbf{0} & \mathbf{0} & \cdots & \widehat{\mathbf{R}}_{\text{SCM},L}\\
		\end{matrix}\right].
	\end{equation}
	This assumption may not hold in practice, but can reduce the computational complexity of the algorithm.  
	
	The authors of \cite{zhao2020beamspace} proposed to construct the spectrum by exploiting the beam-space (BS) of the covariance matrix and the estimated spectrum is given by 
	\begin{equation}\label{BSspec}
		P_{\text{BS},i}
		=\frac{1}{\det\left(\mathbf{A}_i^H \left(\sum\limits_{j=KL+1}^{LN_R}\mathbf{u}_j\mathbf{u}_j^H \right)\mathbf{A}_i  \right)},  i=1,...,N_G
	\end{equation}
	where $\mathbf{u}_j$ denotes the $j$-th eigen-vector of the covariance matrix 
	\begin{equation}
		\label{conSCM}
		\widehat{\mathbf{R}}_{\text{SCM}}=\frac{1}{N_s}\sum_{n=1}^{N_s} \mathbf{y}(n)\mathbf{y}^H(n).
	\end{equation}
	
	However, when the number of samples is small, the estimation accuracy of $\widehat{\mathbf{R}}_{\text{B-SCM}}$ and $\widehat{\mathbf{R}}_{\text{SCM}}$ will be low. Therefore, the spectrum constructed based on $\widehat{\mathbf{R}}_{\text{B-SCM}}$ and $\widehat{\mathbf{R}}_{\text{SCM}}$ will not be accurate, causing bad localization performance. In the following, we propose the ISR method which achieves better localization performance with small number of samples even in the low SNR regime. 
	
	\section{Iterative Sparse Recovery}
	Given the number of targets $K$ is usually much smaller than the number of searching points $N_G$, we propose an iterative localization algorithm by exploiting the sparse structure. Like most direct localization algorithms, the proposed method will estimate a power spectrum for the grid, based on which localization is performed. However, this is done in an iterative manner.   
	
	\subsection{Maximum Likelihood Estimation for the Power Spectrum}
	With the searching grid $\{\bar{\mathbf{p}}_i\}_{i=1}^{N_G}$, (\ref{signalmodel2}) can be reformulated as  
	\begin{equation}\label{signalmodel4}
		\mathbf{y}(n)=\sum_{i=1}^{N_G}\mathbf{A}_i\bar{\mathbf{x}}_i(n)+\mathbf{v}(n),
	\end{equation}
	where $\mathbf{A}_i$ denotes the steering matrix from $L$ SNs to a given searching point $\bar{\mathbf{p}}_i$ as defined in (\ref{ak}), and
	\begin{equation}
		\bar{\mathbf{x}}_i(n)=\left\{\begin{matrix}
			\mathbf{x}_k(n),& \bar{\mathbf{p}}_i = \mathbf{p}_k,\\
			\mathbf{0}, &\bar{\mathbf{p}}_i \neq \mathbf{p}_k.\\
		\end{matrix}\right.
	\end{equation}
	Note that (\ref{signalmodel4}) utilized the sparse structure. 
	
	Different from the existing methods, the ISR algorithm directly utilizes the received signal power from $\bar{\mathbf{p}}_i$ to build the power spectrum, i.e.,
	\begin{equation}
		P_{i}\triangleq \frac{1}{N_R} 	\mathrm{Tr}\left(\mathbf{\Lambda}_i \right).
	\end{equation}
	However, $\mathbf{\Lambda}_i$ is unknown, and the maximum likelihood estimation (MLE) of $\mathbf{\Lambda}_i$ is given by
	\begin{equation}\label{sigmaest}
		\widehat{\mathbf{\Lambda}}_i =\frac{1}{N_s}\sum_{n=1}^{N_s} \bar{\mathbf{x}}_i(n)\bar{\mathbf{x}}_i^H(n), 
	\end{equation}
	for which we need to estimate $\bar{\mathbf{x}}_i(n)$. 
	
	\subsection{Maximum Likelihood Estimation for $\bar{\mathbf{x}}_i(n)$}
	Assuming $\bar{\mathbf{p}}_i$ is the target of interest (TOI), we rewrite (\ref{signalmodel1}) as
	\begin{equation}\label{signalmodel3}
		\mathbf{y}(n)=\mathbf{A}_i\bar{\mathbf{x}}_i(n)+\mathbf{i}(n)+\mathbf{v}(n),
	\end{equation}
	where $\mathbf{i}(n)=\sum_{j\neq i}\mathbf{A}_j\bar{\mathbf{x}}_j(n)$ denotes the interference from other searching points. 
	Then, the covariance matrix for the interference and noise is given as
	\begin{equation}
		\label{incov}
		\begin{aligned}
			\mathbf{R}_{IN,i}&=\mathbb{E}\left(\mathbf{i}(n)\mathbf{i}^H(n)\right)=\mathbf{R}-\mathbf{A}_i\mathbf{\Lambda}_i \mathbf{A}_i^H.
		\end{aligned}
	\end{equation}
	The MLE of $\bar{\mathbf{x}}_i(n)$ can be constructed as a weighted least squares (WLS) problem \cite{li1996adaptive}
	\begin{equation}\label{WLS}
		\hat{\mathbf{x}}_i(n)
		%=\arg \min_{\mathbf{x}} \Vert \mathbf{y}(n)- \mathbf{A}_i\mathbf{x} \Vert_{\mathbf{R}_{IN,i}^{-1}}^2,
		=\arg \min_{\mathbf{x}} \left( \mathbf{y}(n)- \mathbf{A}_i\mathbf{x} \right)^H \mathbf{R}_{IN,i}^{-1} \left( \mathbf{y}(n)- \mathbf{A}_i\mathbf{x} \right),
	\end{equation}
	whose solution is given as
	\begin{equation}\label{WLSsolution}
		\hat{\mathbf{x}}_i(n)=\left(\mathbf{A}_i^H \mathbf{R}_{IN,i}^{-1} \mathbf{A}_i\right)^{-1} \mathbf{A}_i^H \mathbf{R}_{IN,i}^{-1}\mathbf{y}(n).
	\end{equation}
	By substituting (\ref{incov}) into (\ref{WLSsolution}) and using the matrix inverse lemma, we have
	\begin{equation}\label{WLSsolution2}
		\begin{aligned}
			\hat{\mathbf{x}}_i(n)&=\left(\mathbf{A}_i^H \mathbf{R}^{-1} \mathbf{A}_i\right)^{-1} \mathbf{A}_i^H \mathbf{R}^{-1}\mathbf{y}(n).
		\end{aligned}
	\end{equation}
	However, (\ref{WLSsolution2}) requires $\mathbf{R}$, which is unknown. 
	
	\subsection{Estimation of $\mathbf{R}$}
	The estimation of $\mathbf{R}$ is given as
	\begin{equation}\label{Rest}
		\widehat{\mathbf{R}}
		=\sum_{i=1}^{N_G}\mathbf{A}_i \widehat{\mathbf{\Lambda}}_i \mathbf{A}_i^H+\sigma_v^2 \mathbf{I},
	\end{equation}
	where we assume the noise power $\sigma_v^2$ is perfectly obtained. Note that $\widehat{\mathbf{\Lambda}}_i$ is the original variable we want to estimate in (\ref{sigmaest}).
	
	\subsection{ISR Algorithm}
	We need $\widehat{\mathbf{\Lambda}}_i$ to estimate the spectrum for localization purposes. It follows from (\ref{sigmaest}) that 
	$\widehat{\mathbf{\Lambda}}_i$ depends on $\bar{\mathbf{x}}_i(n)$.
	Then, $\hat{\mathbf{x}}_i(n)$ depends on $\mathbf{R}$ as shown in (\ref{WLSsolution2}). Finally, (\ref{Rest}) gives the relation between $\widehat{\mathbf{R}}$ and $\widehat{\mathbf{\Lambda}}_i$.  Therefore, we propose an iterative algorithm to find a fixed point for $\hat{\mathbf{x}}_i(n)$, $\widehat{\mathbf{\Lambda}}_i$, and $\widehat{\mathbf{R}}$ by updating them in a cyclic order, i.e., \begin{equation}\label{iter}
		\cdots \to \widehat{\mathbf{R}}^{(t)} \to \left\{\hat{\mathbf{x}}_i^{(t+1)}(n)\right\}_{n=1}^{N_s} \to \widehat{\mathbf{\Lambda}}_i^{(t+1)} \to \widehat{\mathbf{R}}^{(t+1)} \to \cdots.
	\end{equation}
	where $\hat{\mathbf{x}}_i^{(t)}(n)$, $\widehat{\mathbf{\Lambda}}_i^{(t)}$, and $\widehat{\mathbf{R}}^{(t)}$ denote $\hat{\mathbf{x}}_i(n)$, $\widehat{\mathbf{\Lambda}}_i$, and $\widehat{\mathbf{R}}$ at the $t$-th iteration, respectively. The initialization of $\widehat{\mathbf{R}}$ is given by the block-diagonal SCM as defined in (\ref{SCM}). The proposed ISR algorithm is summarized in Algorithm \ref{algxx}. 
	
	\begin{algorithm}[t]
		\caption{The proposed ISR algorithm} 
		\label{algxx} 
		Input: The samples $\{\mathbf{y}(n)\}_{n=1}^{N_s}$, the searching grid $\left\{\bar{\mathbf{p}}_1,\bar{\mathbf{p}}_2,\cdots,\bar{\mathbf{p}}_{N_G}\right\}$, the noise power $\sigma_v^2$.
		
		Initialize: $t=0$, $\widehat{\mathbf{R}}^{(0)}$ based on (\ref{SCM}).

		Repeat: 
		\begin{enumerate}
			\item Update $\hat{\mathbf{x}}_i^{(t+1)}(n),n=1,\cdots,N_s,i=1,\cdots,N_G,$ based on (\ref{WLSsolution2}), i.e.,
			\[
			\hat{\mathbf{x}}_i^{(t+1)}(n)=\left(\mathbf{A}_i^H \left(\widehat{\mathbf{R}}^{(t)}\right)^{-1} \mathbf{A}_i\right)^{-1} \mathbf{A}_i^H \left(\widehat{\mathbf{R}}^{(t)}\right)^{-1}\mathbf{y}(n).
			\]
			\item Update $\widehat{\mathbf{\Lambda}}_i^{(t+1)}$ based on (\ref{sigmaest}), i.e.,
			% by replacing $\mathbf{x}_k(n)$ with $\hat{\mathbf{x}}_i^{(t+1)}(n)$.
			\[
			\widehat{\mathbf{\Lambda}}_i^{(t+1)} =\frac{1}{N_s}\sum_{n=1}^{N_s} \hat{\mathbf{x}}_i^{(t+1)}(n)\left(\hat{\mathbf{x}}_i^{(t+1)}(n)\right)^H.
			\]
			\item Update $\widehat{\mathbf{R}}^{(t+1)}$ based on (\ref{Rest}), i.e.,
			\[
			\widehat{\mathbf{R}}^{(t+1)}%=\frac{1}{N_s} \sum_{n=1}^{N_s}\mathbf{y}(n) \mathbf{y}^H(n)
			=\sum_{i=1}^{N_G}\mathbf{A}_i \widehat{\mathbf{\Lambda}}_i^{(t+1)} \mathbf{A}_i^H+\sigma_v^2 \mathbf{I}.
			\]
			\item Update the spectrum \[P^{(t+1)}(\hat{\mathbf{p}}_i)=\frac{1}{N_R}\mathrm{Tr}\left(\widehat{\mathbf{\Lambda}}_i^{(t+1)}\right).\]
			\item $t\gets t+1$.
		\end{enumerate}
		Until some termination condition is met.
	\end{algorithm}
	
	\textbf{Remark:}
	We now compare the ISR algorithm with the MVDR (\ref{MVDRspec}) and BS (\ref{BSspec}) based methods. 
	\begin{enumerate}
		\item The MVDR method assumes the block-diagonal covariance structure in (\ref{SCM}), which reduces the computational complexity at the cost of worse performance. Furthermore, it estimates the received signal power from one searching point based on the output of the MVDR combiner, which may not be optimal.
		
		\item The BS method estimates $\mathbf{R}$ by the conventional SCM method (\ref{conSCM}), which is theoretically optimal but requires a large number of samples. As a result, the estimation performance will suffer from limited number of samples. 
		
		\item The proposed IRS method utilizes sparse recovery technique to 
		represent the received signal in a lower dimension and reduces the searching space. Furthermore, by connecting three key parameters, $\hat{\mathbf{x}}_i(n)$, $\widehat{\mathbf{\Lambda}}_i$, and $\widehat{\mathbf{R}}$, the proposed method iteratively improves the estimation accuracy even with very few samples, achieving a communication-efficient localization.
	\end{enumerate}
	
	\begin{figure}[!t]
		\centering
		\includegraphics[width=2.3in]{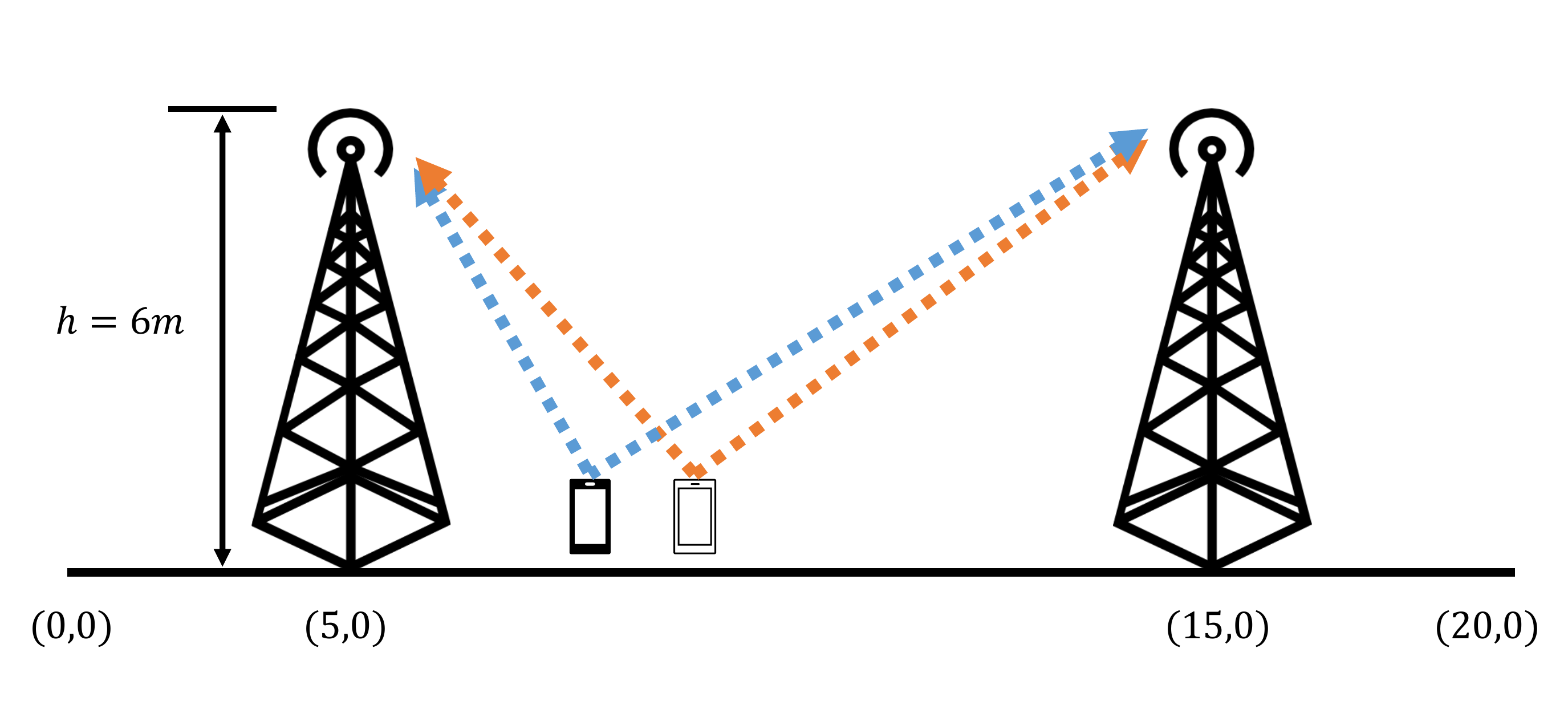}
		\caption{Illustration of scenario 1.}
		\label{fig_Simu1_illu}
	\end{figure}
	
	\section{Simulation}
	In this section, we demonstrate the performance of the proposed  algorithm by simulation. We consider an OFDMA systems where the transmitted signals from different targets are assumed to be single-carrier signals with different frequencies. The signal-to-noise ratio (SNR) is defined as $\mathrm{SNR}=\frac{1}{LK}\sum_{l=1}^{L}\sum_{k=1}^{K}\frac{\sigma_{l,k}^2}{\sigma_v^2}.$
	For comparison purpose, we consider two scenarios. 
	
	\subsection{Scenario 1: Linear Monitoring Area}
	We begin with a simple case where two SNs and two targets are located on the same line, as illustrated in Fig. \ref{fig_Simu1_illu}. 
	Assume the two SNs are located at $(5,0)$ and $(15,0)$, where all coordinates are in meters. The monitoring area is the line from $(0,0)$ to $(20,0)$ with step 0.1. The height of the received antenna arrays is 6 meters and each SN is equipped with $N_R=64$ antennas. The users transmit OFDM signals on a typical 802.11ac WiFi standard with $\Delta f = 312.5 kHz$ and $|s_k(n)|=1$.
	
	\begin{figure}[h]
		\centering
		\includegraphics[width=2.7in]{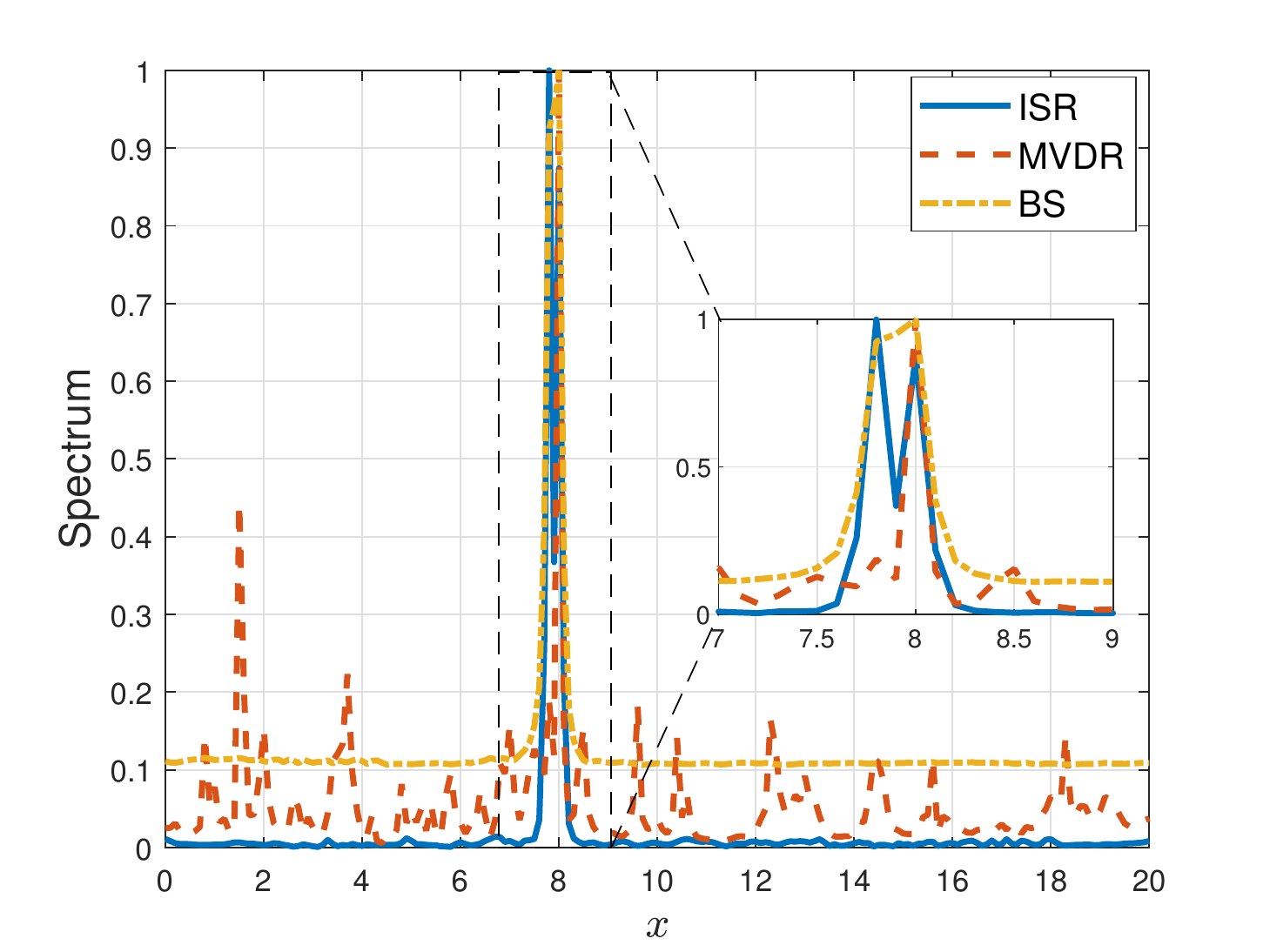}
		\caption{Comparison of the power spectrum with 2 targets.}
		%under different positions of targets.}
	\label{fig_Simu1_spec}
\end{figure}

Fig. \ref{fig_Simu1_spec} shows the estimated power spectrum for the two targets located at $(7.8,0)$ and $(8.0,0)$, which are close to each other. The SNR is set as $-5$ dB and the number of samples is $N_s=2$. It can be observed from Fig. \ref{fig_Simu1_spec} that the ISR algorithm achieves better resolution than the MVDR and BS methods. In particular, the MVDR-based method only has one peak corresponding to the target at $(8,0)$, while the BS method can not differentiate the two targets. %\xl{How many samples here?}

\begin{figure}[!t]
	\centering
	\includegraphics[width=2.7in]{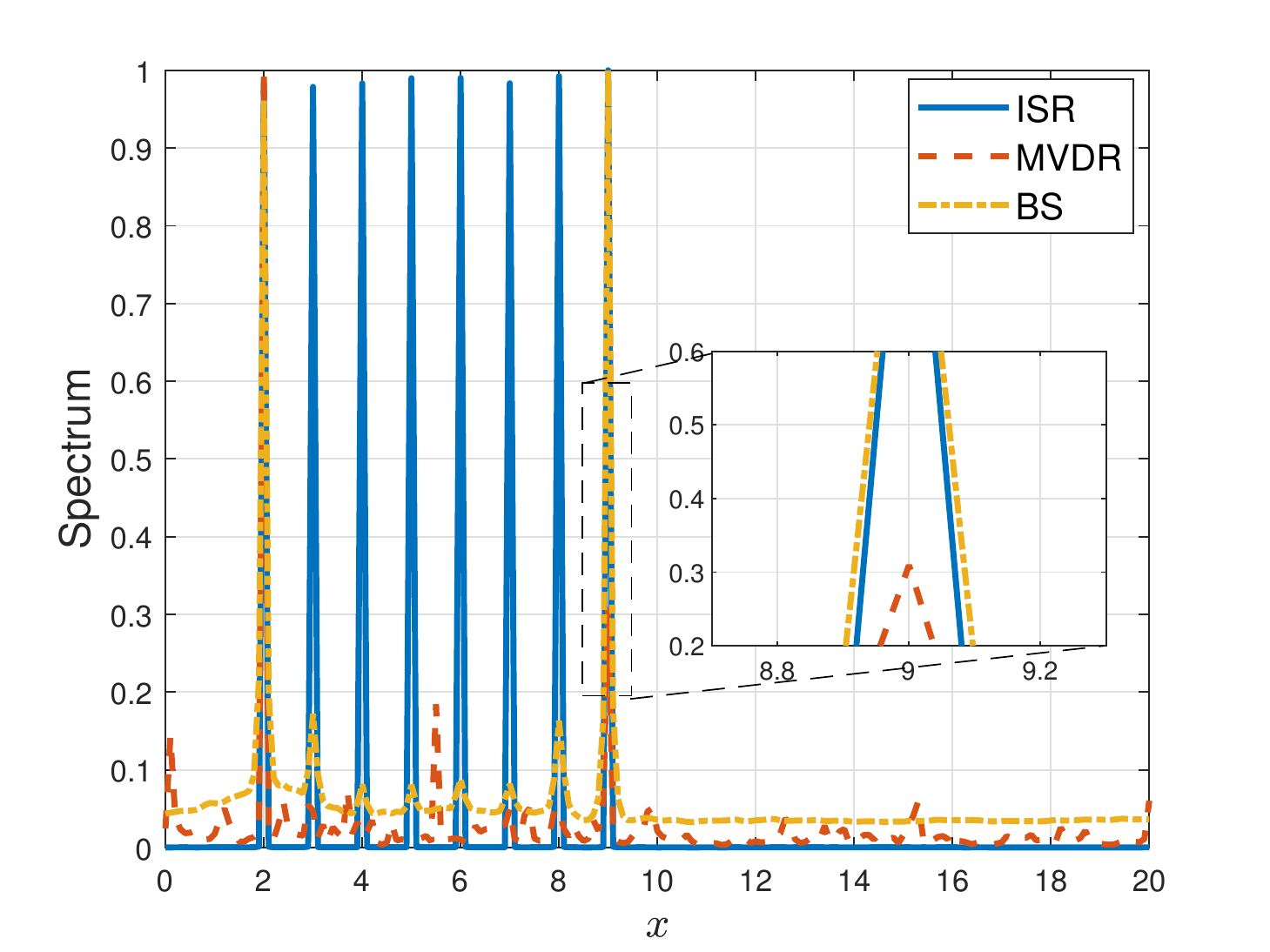}
	\caption{Comparison of the power spectrum with 8 targets.}
	\label{fig_Simu1_spec2}
\end{figure}

Fig. \ref{fig_Simu1_spec2} shows the power spectrum for eight targets uniformly distributed between $(2,0)$ and $(9,0)$. The SNR is set as $\mathrm{SNR}=0$ dB and the number of samples is $N_s=8$. We can observe from Fig. \ref{fig_Simu1_spec2} that the ISR algorithm achieves good localization performance, while the MVDR and BS methods are not able to differentiate all targets. This is because the number of samples is not sufficient for the MVDR and BS methods to recover the subspace of all targets from the SCM.

\begin{figure}[!t]
	\centering
	\includegraphics[width=2.7in]{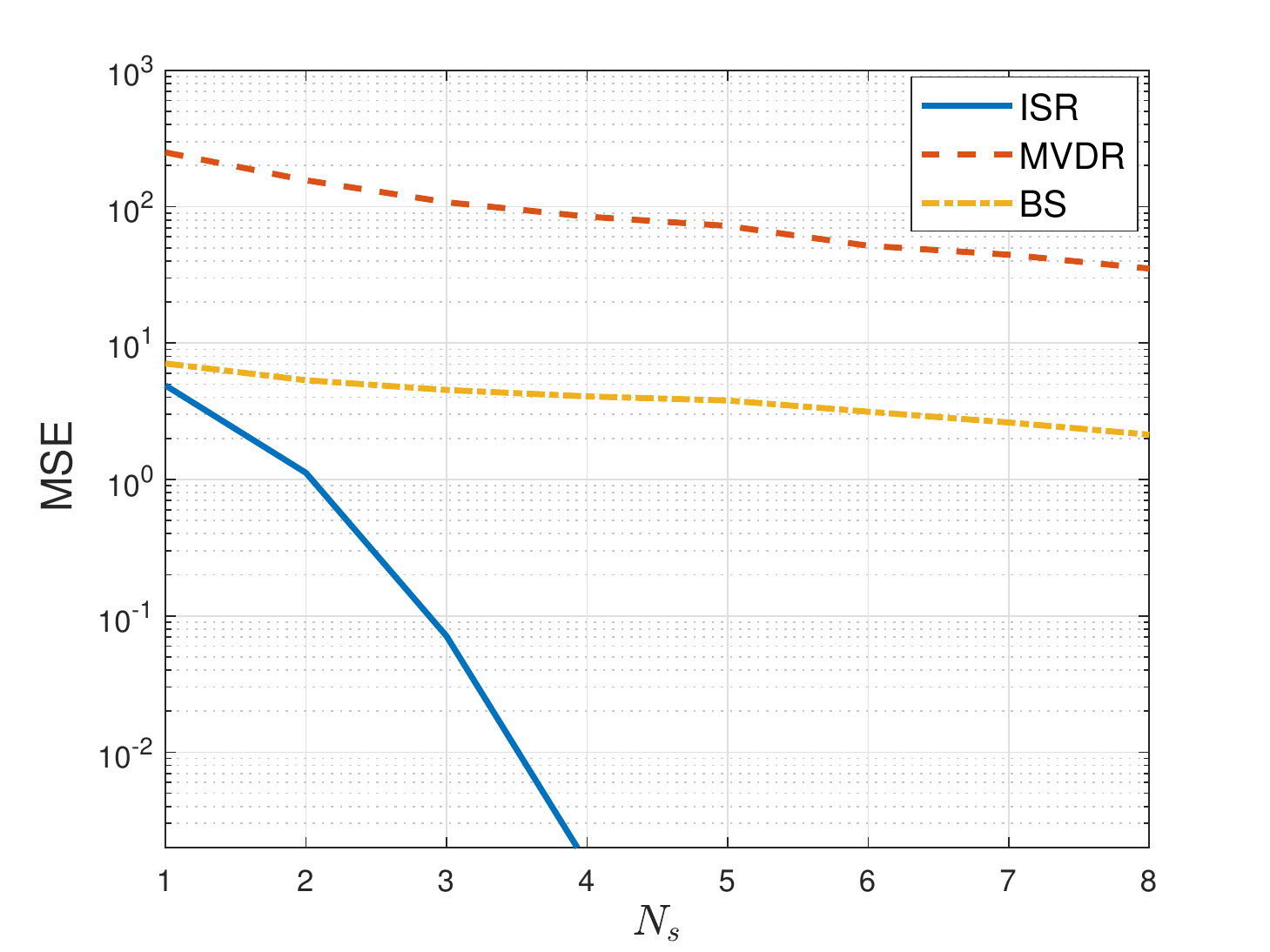}
	\caption{Comparison of the MSE versus the number of samples. }
	\label{fig_Simu1_mse}
\end{figure}

To quantify the estimation accuracy, we utilize the mean squared error (MSE), i.e., $	\text{MSE}=\frac{1}{N_{mc}}\sum_{n=1}^{N_{mc}} \left\Vert\hat{\mathbf{p}}_n-\mathbf{p}\right\Vert^2$
as the performance metric, where $\hat{\mathbf{p}}_n$ denotes the estimated position at the $n$-th Monte-Carlo trial and $N_{mc}$ represents the number of Monte-Carlo trials. Here we set $N_R=36$, $\text{SNR}=0$dB, $N_{mc}=10^4$ and consider four targets, which are uniformly located from $(2,0)$ to $(5,0)$. The localization is achieved by finding the highest peak in the power spectrum.
Fig. \ref{fig_Simu1_mse} compares the MSE of different methods versus the number of samples. It can be observed that the ISR algorithm can achieve better accuracy than the other two methods, especially when $N_s\geq 2$. How to improve the performance with $N_s=1$ would be left as future work.

\subsection{Scenario 2: Square Monitoring Area}

\begin{figure}[!t]
	\centering
	\includegraphics[width=2.3in]{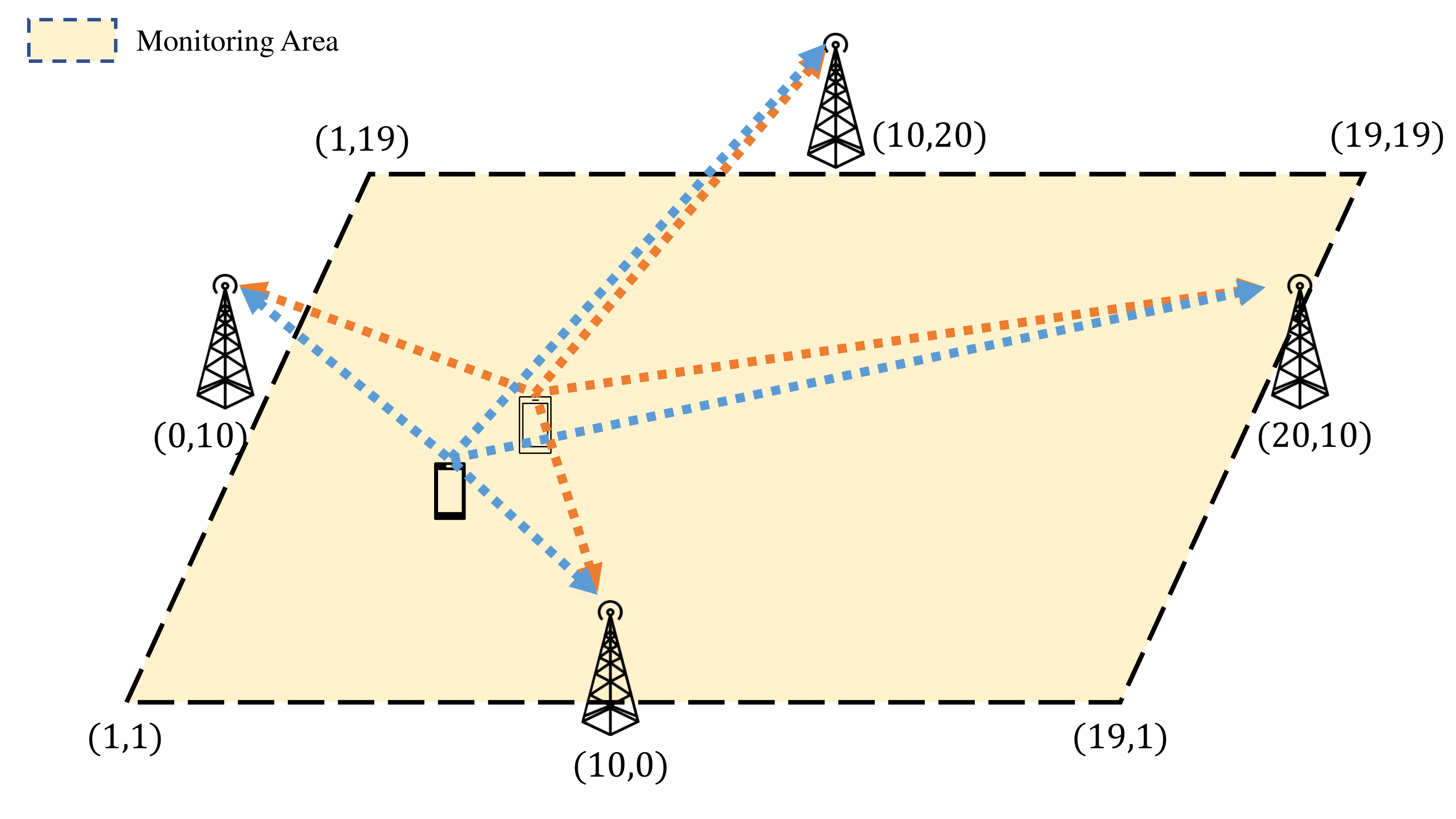}
	\caption{Illustration of scenario 2.}
	\label{fig_Simu2_illu}
\end{figure}

Next, we consider a square monitoring area with 4 SNs, as illustrated in Fig. \ref{fig_Simu2_illu}. The SNs are placed at $(0,10)$, $(10,20)$, $(20,10)$, and $(10,0)$, and equipped with 64 antennas. The searching area is from $1$ to $19$ with step $0.5$ in both $x$ and $y$ directions. Two targets are placed at $(3.5,13.5)$ and $(3.5,14.5)$ and the SNR is set as $3$ dB. 

\begin{figure*}[h]
	\centering
	\subfloat[]{\includegraphics[width=2.3in]{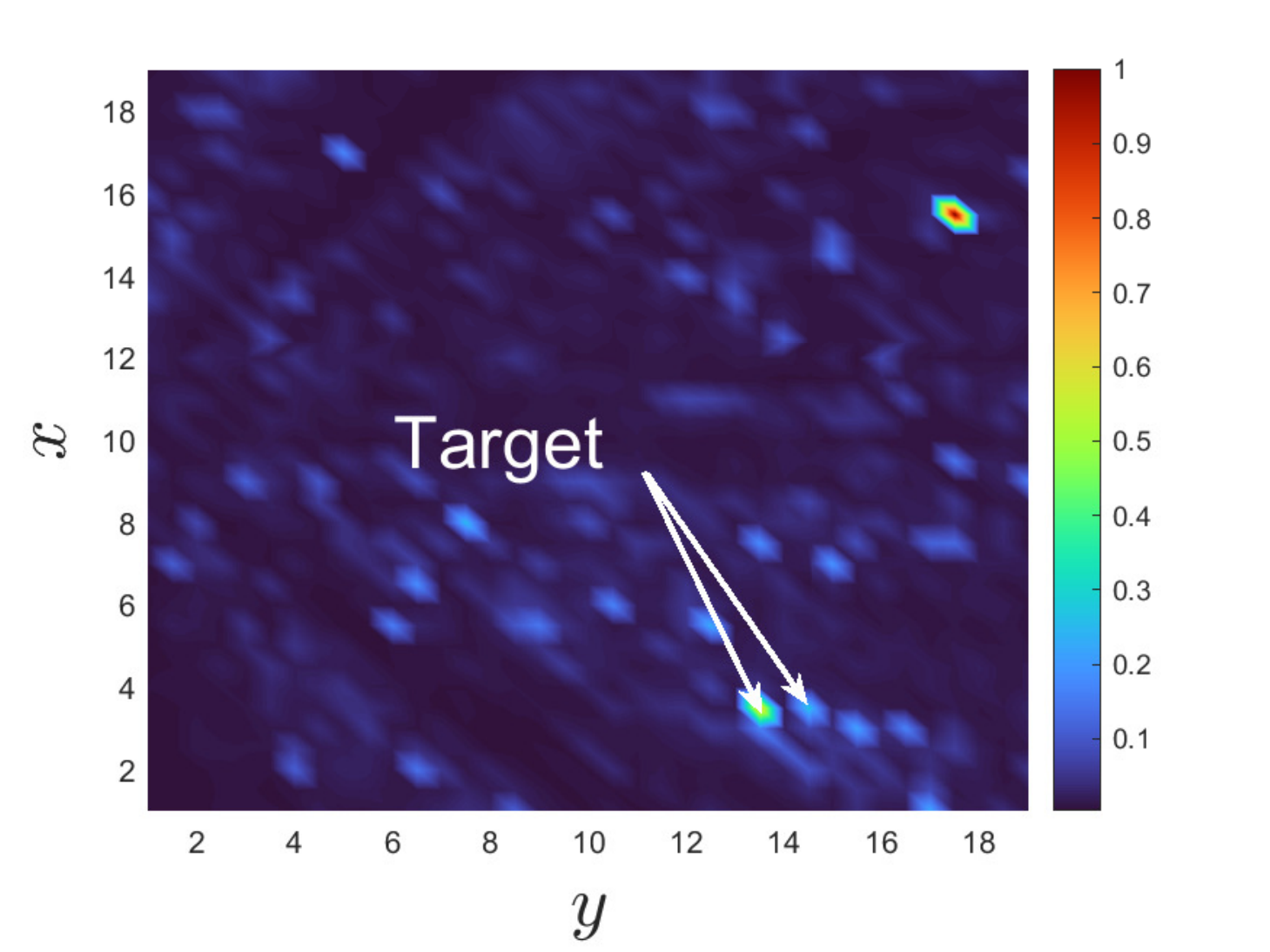}}\
	\subfloat[]{\includegraphics[width=2.3in]{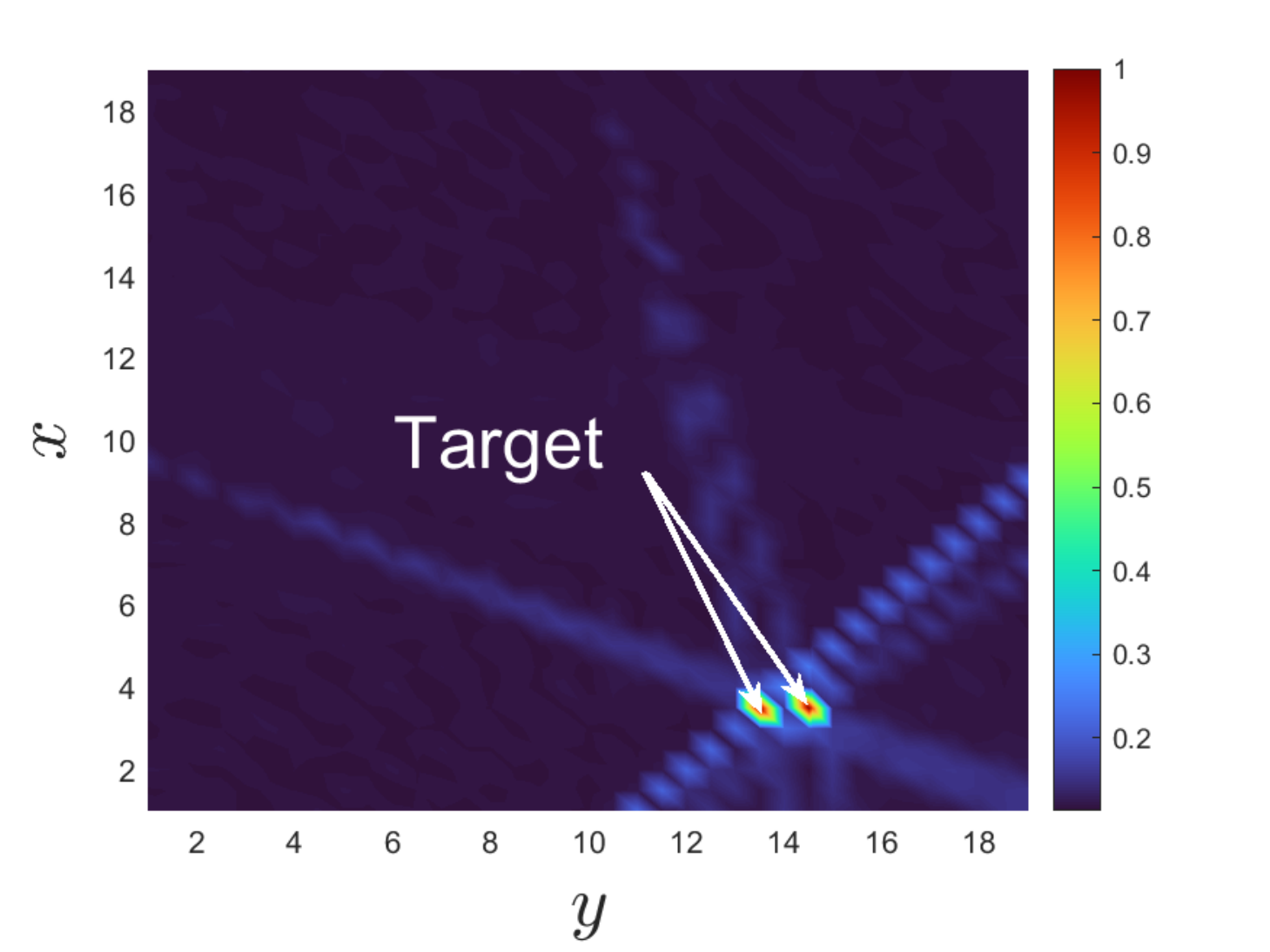}}\
	\subfloat[]{\includegraphics[width=2.3in]{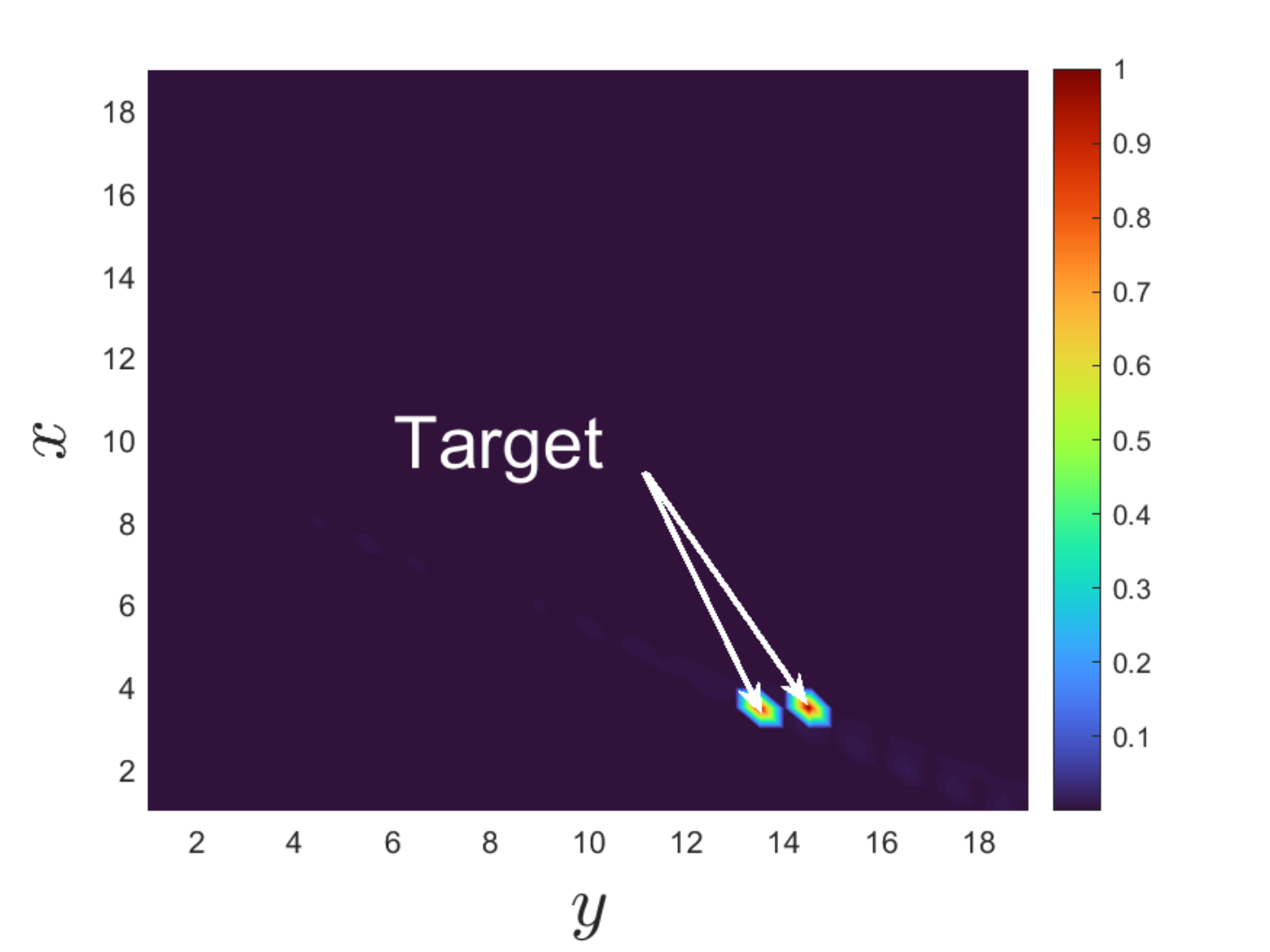}}\
	\caption{Comparison of the power spectrum with 2 targets. (a) MVDR; (b) BS; (c) ISR.}
	\label{fig_Simu2_spec}
\end{figure*}

Fig. \ref{fig_Simu2_spec} shows the power spectrum estimated by three different methods with $N_s=2$. It can be observed that there are peaks at the positions corresponding the actual targets for all three methods. However, the peaks for MVDR are week and will be submerged by other false peaks. The BS method achieves better performance than MVDR, but the ISR algorithm can clearly form two peaks in the power spectrum with a clean background.

\begin{figure*}[h]
	\centering
	\subfloat[]{\includegraphics[width=2.3in]{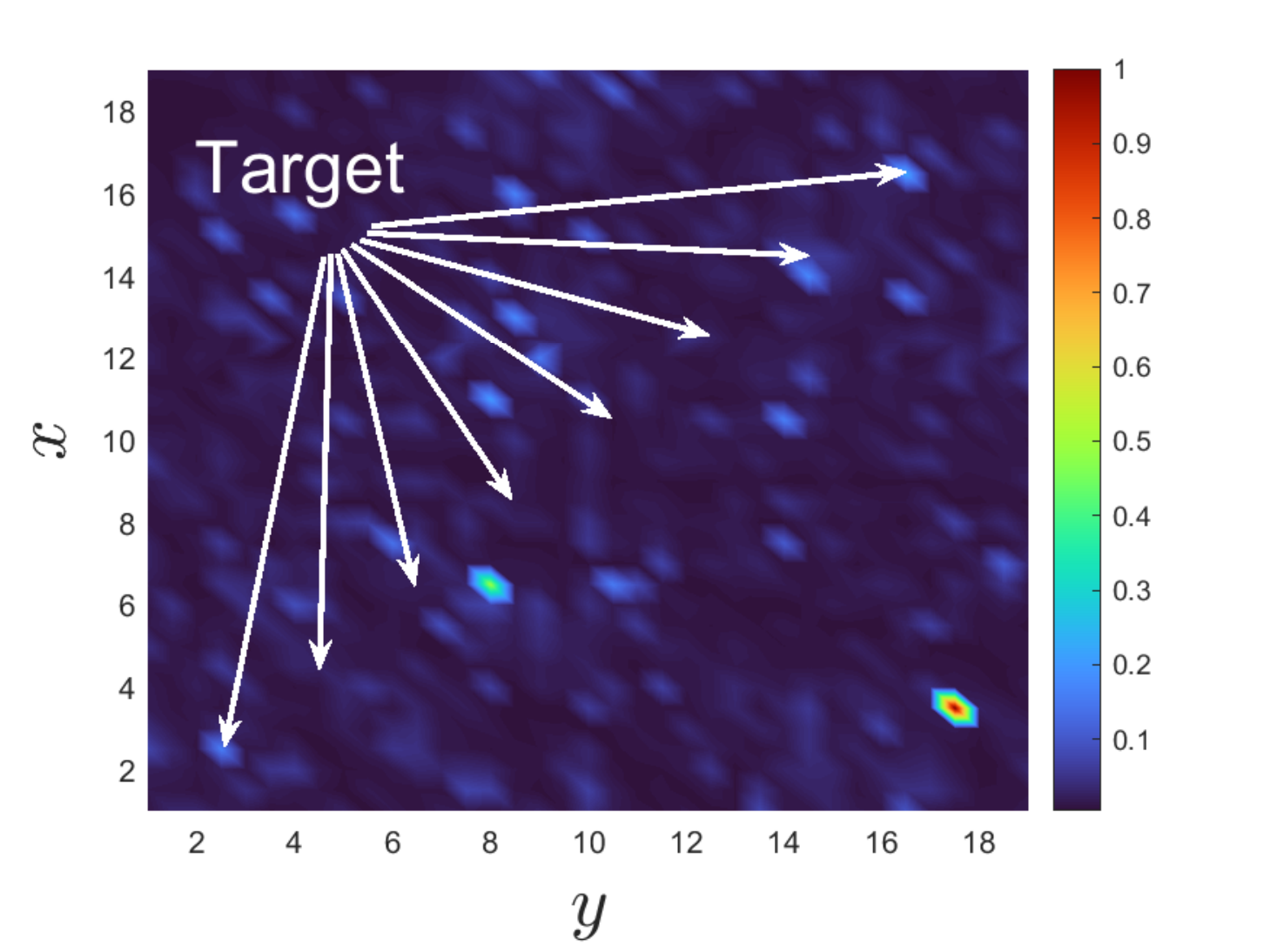}}\
	\subfloat[]{\includegraphics[width=2.3in]{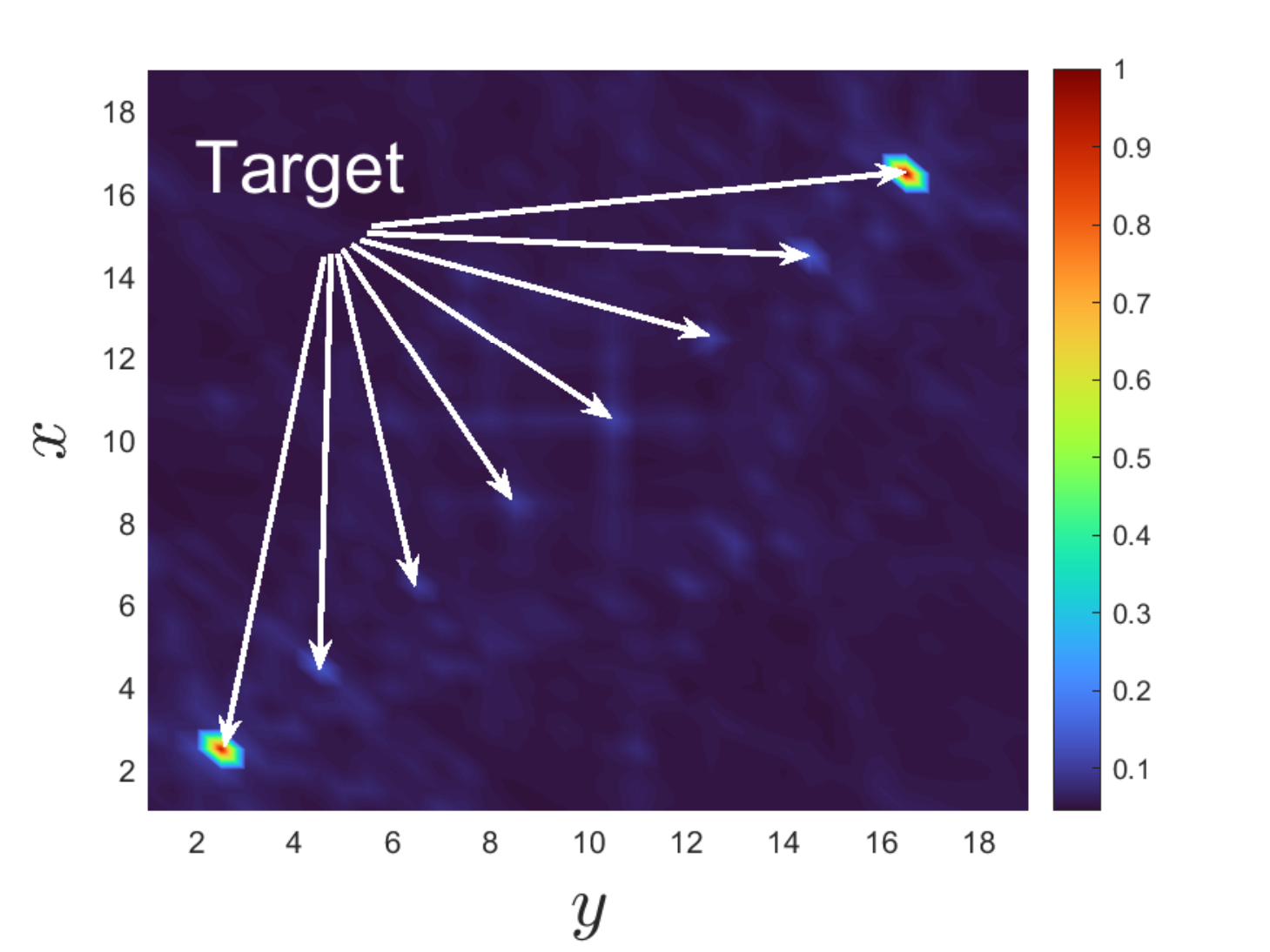}}\
	\subfloat[]{\includegraphics[width=2.3in]{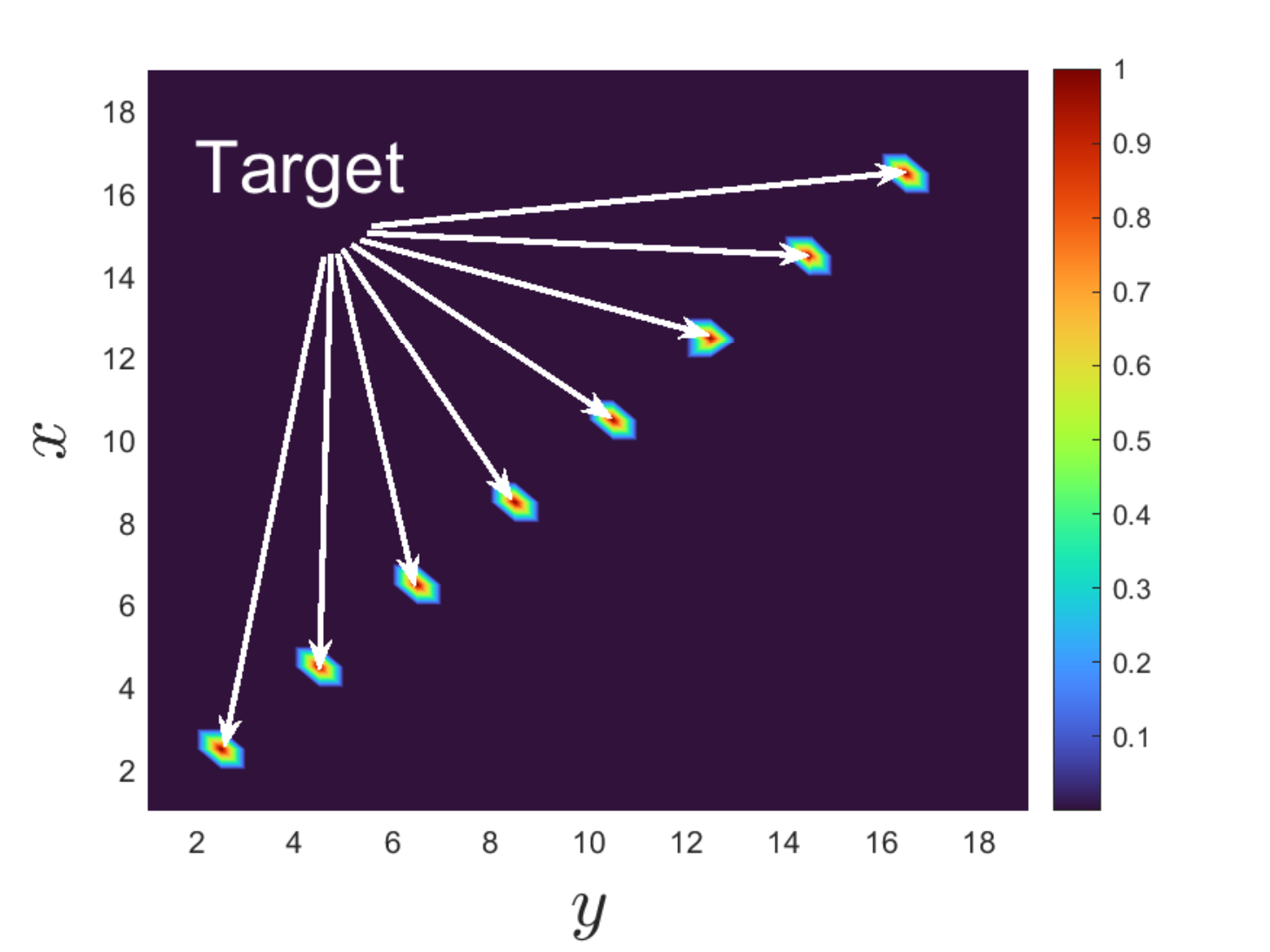}}\
	\caption{Comparison of the power spectrum with 8 targets. (a) MVDR; (b) BS; (c) ISR.}
	\label{fig_Simu2_spec2}
\end{figure*}

Finally, we consider a more complex scenario with 8 targets, which are uniformly located from $(2.5,2.5)$ to $(16.5,16.5)$. The SNR is set as $5$ dB and $N_s=16$. We can observe from Fig. \ref{fig_Simu2_spec2} that some targets can not be detected by the MVDR and BS methods, whereas the ISR algorithm can still provide good localization performance.

\section{Conclusion}
This paper considered the direct passive localization in perceptive mobile networks, where the localization of several targets is collaboratively achieved by several sensing nodes based on the uplink communication signals. Different from existing methods, the proposed ISR algorithm exploited the sparse structure of the problem and improved the estimation accuracy for the covariance matrix and the power spectrum in an iterative manner. Experiment results showed the better localization performance of the proposed ISR algorithm than existing methods. The work in this paper demonstrated the advantage of the ISR algorithm and provided a communication-efficient multi-target localization method for future perceptive mobile networks. 

\section*{Acknowledgements}
We are grateful to the anonymous reviewers for their valuable and constructive comments. This work was supported by the HKUST-BDR Joint Research Institute under Grant OKT22EG04.

\end{document}